\documentclass[aps,preprint,showpacs]{revtex4}
\usepackage{graphicx}

\begin{document}

\title{Effective moment of inertia for several fission reaction systems induced by nucleons, light particles and heavy ions}

\author{S. Soheyli\footnote{Corresponding author: soh@basu.ac.ir}}
\affiliation{Bu-Ali Sina University, Department of Physics, Hamedan,
 Iran}
\begin{abstract}
Compound nucleus effective moment of inertia has been calculated for
several fission reaction systems induced by nucleons, light
particles, and heavy ions. Determination of this quantity for these
systems is based upon the comparison between the experimental data
of the fission fragment angular distributions as well as the
prediction of the standard saddle-point statistical model (SSPSM).
For the systems, the two cases, namely with and without neutron
emission corrections were considered. In these calculations, it is
assumed that all the neutrons are emitted before reaching the saddle
point.It should be noted that the above method for determining of
the effective moment of inertia had not been reported until now and
this method is used for the first time to determine compound nucleus
effective moment of inertia. Hence, our calculations are of
particular importance in obtaining this quantity, and have a
significant rule in the field of fission physics. Afterwards, our
theoretical results have been compared with the data obtained from
the rotational liquid drop model as well as the Sierk model, and
satisfactory agreements were found. Finally, we have considered the
effective moment of inertia of compound nuclei for the systems that
formed similar compound nuclei at similar excitation energies.
\end{abstract}

 \maketitle

The angular distribution of fission fragments is an effective probe
to study the dynamics of the fission reaction. The effective moment
of inertia of the compound nucleus is a crucial quantity in
determining fusion-fission dynamics, although there has not been
introduced any precise method to determine it. In this paper, to
calculate the effective moment of inertia of the compound nucleus, a
novel method is presented. In this method, the values of the
effective moment of inertia of the compound nucleus for several
induced-fission reaction systems  by light projectiles and heavy
ions are determined by fission fragment angular distribution method.
This method is based upon the comparison between the experimental
data of the fission fragment angular distribution as well as the
prediction of the standard saddle point statistical model(SSPSM).
Calculation through this method is considered with and without
neutron emission correction in the reactions. It is now clear that
because of the hindrance to fission, a larger number of particles,
neutrons in particular, are emitted from the fissioning system.
These pre-fission neutrons are emitted not only during the
transition stage up to the saddle point but also during the decent
from the saddle to the scission point. The emission of neutrons
before reaching the saddle point has the effect of lowering the
available excitation energy at this point and this in turn will
reduce the variance of the $K$ distribution. Assuming the average
angular momentum removed as 0.5 per neutron, correction to $<I^2>$
values can also be made. However, the change in the variance of the
$K$ distribution due to the angular momentum of emitted neutrons is
not significant. In the present work, the pre-fission neutrons are
taken to be emitted before the saddle point, since it is not
straightforward to separate experimentally the contribution of
neutrons emitted before the saddle point and the ones emitted after
the saddle point but before the scission point. Afterwards, our
theoretical results have been compared with the data obtained from
the rotational liquid drop model as well as the rotating finite
range model(RFRM), and satisfactory agreements were found. Finally,
we have compared the effective moment of inertia for the systems,
populating the same compound nucleus and at similar excitation
energies.

A study of fission process will produce its own unique insights into
the problem of large-scale nuclear dynamics. The study of the
fission fragment angular distribution of induced fission by light
projectiles and heavy ions continues to be a rich source of
information.  There is an important physical quantity in the
calculation of angular anisotropy of fission fragment by statistical
models. This quantity is the effective moment of inertia of the
compound nucleus.  Fission fragment anisotropy $A$ defined as: $A=
\frac{W(0^o or 180 ^o)}{W(90^o)}$. According to the SSPSM,
anisotropy $A$ that is proportional to $<I^2>$ is given by: $A \cong
1 + \frac{< I^2
>\hbar^2}{4 \Im_{eff}T_{sad}}$, where $< I^2 >$ is the mean-square angular momentum of the compound
nucleus and
$\Im_{eff}=\frac{\Im_{\parallel}\Im_{\perp}}{(\Im_{\perp}-\Im_{\parallel})}$
is the effective moment of inertia of the compound nucleus at the
saddle point; $\Im_{\parallel}$ and $\Im_{\perp}$ are moments of
inertia parallel and perpendicular to the symmetry axis,
respectively \cite{vandenbosh:1973}. $T_{sad}$,  the nuclear
temperature at the saddle point can be calculated as:
$T_{sad}=\sqrt{\frac{E_{ex}}{\emph{a}}}=\sqrt{\frac{E_{c.m.}+Q-B_{f}-E_{R}-{\nu}E_{n}}{\emph{a}}}.
\label{nucl. tem}$ In this equation,  $E_{ex}$  denotes the
excitation energy of the compound nucleus, while $E_{c.m.}$, $Q$,
$B_{f}$, $E_{R}$, $\nu$ and $E_{n}$ represent the center-of-mass
energy of the projectile, the $Q$ value, fission barrier height, ,
rotational energy of the compound nucleus, the number of pre-fission
neutrons, and the average energy of an emitted neutron,
respectively. The quantity $\emph{a}$~ stands for the level density
parameter at the saddle point.
 Due to the use of the SSPSM, in the calculations,
it is assumed that neutrons are emitted before the compound nucleus
approaches the saddle point. The average energy carried by an
emitted neutron in fission reactions systems induced by nucleons and
light particles is assumed to be 5 MeV as well as this energy for
each emitted neutron is about 9-10 MeV in induced fissions by heavy
ions.
 The calculation method is based upon the experimental data for the
fission fragment angular distribution. It is found that $\Im_{eff}$
is dependent on the second moment of the compound nucleus spin
distribution. Afterward, the quantity $\Im_{eff}$ is considered as a
linear equation in terms of the center-of-mass energy of the
projectile as $\Im_{eff}=a_{\nu}E_{c.m.}+b_{\nu}$, where $\nu=0, 1,
2$ denotes considering without the emission of neutron, emission of
one neutron, emission of two neutrons, respectively.

 The coefficients of equation $\Im_{eff}$ in terms of $E_{c.m.}$ without considering
neutron emission and by considering the correction related to
neutron emission for 12 systems undergoing induced fission with
nucleons and light particles are given in Table I. The references
for finding all necessary data including $B_{f}$, $<I^{2}>$, and the
experimental angular distribution data  and the bombarding energy
range for the systems are also listed in Table I.

\begin{table}[p]
\begin{center}
\renewcommand{\arraystretch}{1.3}
\begin{tabular}{c  c c c c c c c c c}
\hline fission systems&~\texttt{C. N.}&~$a_{\circ}~$& ~$b_{\circ}$~&~$a_{1}$~&~$b_{1}$~&~$a_{2}$~&~$ b_{2}$&$E_{proj.}(MeV)$&~References\\
\hline \hline
${\texttt{P}+^{185}\texttt{Re}}$&$^{186}\texttt{Os}$& 0.362&~~ 22.291&~~ 0.284&~~ 29.890&~~ 0.144&~~ 41.925&42-66.5&[2, 4]\\
${\tau+^{183}\texttt{W}}$&$^{186}\texttt{Os}$& 0.421&~~ 31.853&~~ 0.254&~~ 43.966&~~ - &~~ 68.466&28-59.5&~~~ [2, 6, 7]\\
${\alpha+^{182}\texttt{W}}$&$^{186}\texttt{Os}$& 0.312&~~ 21.005&~~ 0.217&~~ 30.483&~~ 0.038&~~ 46.420&53.5-73&[2, 4]\\
${\alpha+^{185}\texttt{Re}}$&$^{189}\texttt{Ir}$& 0.398&~~ 16.573&~~ 0.296&~~ 26.367&~~ 0.087&~~ 44.156&46-72.5&[2, 4]\\
${\texttt{P}+^{197}\texttt{Au}}$&$^{198}\texttt{Hg}$& 0.412&~~ 20.832&~~ 0.327&~~ 28.711&~~ 0.162&~~ 41.883&36-66&[2, 4]\\
${\tau+^{197}\texttt{Au}}$&$^{200}\texttt{Tl}$& 0.366&~~ 50.332&~~ 0.096&~~ 69.203&~~ - &~~ 124.503&23-59&~~~ [2, 6, 7]\\
${\alpha+^{197}\texttt{Au}}$&$^{201}\texttt{Tl}$& 0.239&~~ 33.965&~~ 0.105&~~ 46.646&~~ - &~~ 68.083&47.5-77&[2, 4]\\
${\texttt{P}+^{205}\texttt{Tl}}$&$^{206}\texttt{Pb}$& 0.165&~~ 36.319&~~ - &~~ 50.080&~~ - &~~ 77.087&36-67.5&[2, 4]\\
${\tau+^{207}\texttt{Pb}}$&$^{210}\texttt{Po}$& 0.791&~~ 36.373&~~ 0.473&~~ 57.958&~~ -&~~ -&21-59&~~~ [2, 6, 7]\\
${\alpha+^{206}\texttt{Pb}}$&$^{210}\texttt{Po}$& 0.104&~~ 39.028&~~ - &~~ 53.639&~~ - &~~ 80.834&45.5-72.5&[2, 3]\\
${\alpha+^{209}\texttt{Bi}}$&$^{213}\texttt{At}$& 0.276&~~ 41.611&~~ 0.057&~~ 60.694&~~ -&~~ 104.841&44-76&[2, 4]\\
${\texttt{n}+^{232}\texttt{Th}}$&$^{233}\texttt{Th}$& 0.478&~~ 19.239&~~ 0.417&~~ 24.934&~~ 0.289&~~ 34.746&21-95&~~~ [2, 5, 8]\\
\hline \hline
\end{tabular}
\caption{\label{eq. table} Coefficients of equation $\Im_{eff}$ in
terms of $E_{c.m.}$ for light particle induced fission systems
without considering neutron emission and by considering correction
related to neutron emission.}
\end{center}
\end{table}

In Table I, for some reaction systems, the effective moment of
inertia of the compound nucleus is calculated  without the emission
of neutron, or by taking at most one emitted neutron into account,
because the coefficients of equation $\Im_{eff}$ in terms of
$E_{c.m.}$ will be negative in terms of emitting two neutrons. It
may be noted that the center-of-mass energy of the projectile is
roughly the same as that in the laboratory framework for the induced
fission performed by light projectiles as well as the rotational
energy $E_{R}$ can be neglected. The effective moment of inertia for
the $\texttt{p}+^{209}\texttt{Bi}$ reaction system has also been
calculated, but the coefficient of $E_{c.m.}$ in equation
$\Im_{eff}$ will be negative considering without the emission of
neutron. The effective moment of inertia for this system reaction is
similar to the ${\tau+^{207}\texttt{Pb}}$ and
${\alpha+^{206}\texttt{Pb}}$ reaction systems at the same excitation
energy for the compound nucleus, since these systems that formed
similar compound nucleus, $^{210}\texttt{Po}$. In this paper, level
density parameter, a is taken 0.094$\texttt{A}_{\texttt{C.N.}}$ for
all the reaction systems, where $\texttt{A}_{\texttt{C.N.}}$ is the
mass number of compound nucleus.

 The coefficients of equation $\Im_{eff}$ in terms of
$E_{c.m.}$ without considering neutron emission and by considering
the correction related to neutron emission for 6 systems undergoing
induced fission with heavy ions are given in Table II. The
references for finding all necessary data including $B_{f}$,
$<I^{2}>$, $E_{R}$ and the experimental angular distribution data
are also given. The bombarding energy range for the systems are also
listed in Table II. For all these reaction systems, a is also taken
$0.094\texttt{A}_{\texttt{C.N.}}$.
\begin{table}[p]
\begin{center}
\renewcommand{\arraystretch}{1.3}
\begin{tabular}{c  c c c c c c c c c}
\hline fission systems&~\texttt{C. N.}&~$a_{\circ}~$& ~$b_{\circ}$~&~$a_{1}$~&~$b_{1}$~&~$a_{2}$~&~$ b_{2}$&$E_{proj.}(MeV)$&~References\\
\hline \hline
${^{12}\texttt{C}+^{232}\texttt{Th}}$&$^{244}\texttt{Cm}$& 1.155&~~ 55.016&~~ 1.033&~~ 72.243&~~ 0.841&~~ 96.353&64-78&~ [9, 10]\\
${^{11}\texttt{B}+^{235}\texttt{U}}$&$^{246}\texttt{Bk}$& 10.637&~~ -538.514&~~ 11.086&~~ -558.763&~~ 11.581&~~ -580.331&60-70&~ [9, 10]\\
${^{14}\texttt{N}+^{232}\texttt{Th}}$&$^{246}\texttt{Bk}$& 4.105&~~ -247.861&~~ 4.236&~~ -254.123&~~ 4.375&~~ -260.208&75-86&~ [9, 10]\\
${^{11}\texttt{B}+^{237}\texttt{Np}}$&$^{248}\texttt{Cf}$& 1.910&~~ 14.855&~~ 1.872&~~ 25.527&~~ 1.813&~~ 39.215&75-114&~ [9, 10]\\
${^{12}\texttt{C}+^{236}\texttt{U}}$&$^{248}\texttt{Cf}$& 1.284&~~ 45.461&~~ 1.243&~~ 56.276&~~ 1.185&~~ 69.816&75-125&~ [9, 10]\\
${^{16}\texttt{O}+^{232}\texttt{Th}}$&$^{248}\texttt{Cf}$& 2.087&~~ -54.563&~~ 2.072&~~ -45.811&~~ 2.042&~~ -34.389&93-136&~ [9, 10]\\
\hline \hline
\end{tabular}
\caption{\label{eq. table} Coefficients of equation $\Im_{eff}$ in
terms of $E_{c.m.}$ for heavy ion induced fission systems without
considering neutron emission and by considering correction related
to neutron emission.}
\end{center}
\end{table}
 The calculated values for $\Im_{eff}$ for the systems undergoing
induced fission with nucleons and light particles are compared with
those of rotational liquid drop model(RLDM)[11]. It was observed the
calculated values to be in agreement with the values obtained by
RLDM. In Fig. 1, the effective moment of inertia for the
${\alpha+^{182}\texttt{W}}$ system are compared with those of the
RLDM as a typical.

\begin{figure}[h]
\centering
\includegraphics[scale=0.7]{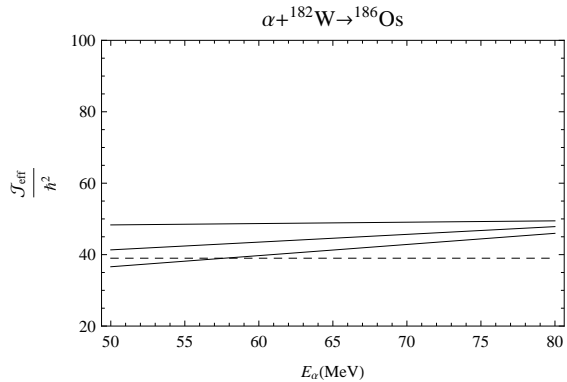}

\caption{ The comparison of $\Im_{eff}$ obtained from fission
fragment angular distribution method with those of RLDM (dashed
line) for the
${\alpha+^{182}\texttt{W}\longrightarrow^{186}\texttt{Os}}$ system.
The solid lines from bottom to top represent the effective moment of
inertia without correction of neutron emission, with neutron
emission correction when the number of emitted neutrons is
considered 1 and 2, respectively.}  \label{Cf}
\end{figure}

In Fig. 2, the $\Im_{eff}$ for the
${^{12}\texttt{C}+^{236}\texttt{U}}$ system are compared with those
of the rotating finite range model(RFRM) by Sierk[12] as a typical.
The calculated effective moment of inertia in this method was seen
to be in agreement with the values obtained by the RFRM.
\begin{figure}[h]
\vspace{1 cm}
\centering
\includegraphics[scale=0.7]{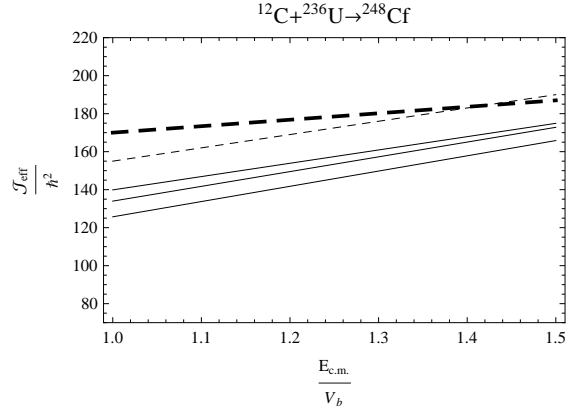}

\caption{ The comparison of $\Im_{eff}$ obtained from fission
fragment angular distribution method with those of RFRM(the
long-dashed line) for the
${^{12}\texttt{C}+^{236}\texttt{U}}\longrightarrow^{248}\texttt{Cf}$
system. The solid lines from bottom to top represent the effective
moment of inertia without correction of neutron emission, with
neutron emission correction when the number of emitted neutrons is
considered 1 and 2, respectively. The dashed line is the calculated
$\Im_{eff}$ for the system by considering correction related to the
average number of emitted neutrons[10].} \label{Bk}
\end{figure}\

  Finally,  we have compared the effective moment of inertia
for the reaction systems that formed similar compound nuclei. It is
expected that this quantity must be the same for the reaction
systems populating the same compound nuclei at the same excitation
energies.  For the ${\texttt{P}+^{185}\texttt{Re}}$,
${\tau+^{183}\texttt{W}}$, and ${\alpha+^{182}\texttt{W}}$ reaction
systems that formed similar compound nucleus $^{186}\texttt{Os}$, it
is observed that the average errors in the calculation of effective
moment of inertia of the compound nucleus $^{186}\texttt{Os}$, over
the energy rang for the reactions mentioned above are $<5\%$ and
$\simeq29\%$ for the former two reactions in comparison the last
reaction, respectively. Keeping in mind that the effective moment of
inertia for the reaction systems populating the same compound
nucleus are independent the role of the entrance channel for these
reaction systems and depends on the excitation energy of the
compound nucleus. We propose here that the discrepancy between the
calculated $\Im_{eff}$ for the ${\alpha+^{182}\texttt{W}}$ reaction
system and with those of the ${\tau+^{183}\texttt{W}}$ system arises
from the measurement technique of fission fragments(catcher foil
technique) which is used in the last experiment[6]. The fission
fragment angular distributions are measured using different types of
detectors both passive like mica, lexan and glass and active like
gas and silicon surface barrier detectors. A combination of gas and
silicon detectors have also been used for a cleaner separation of
fission fragments.

  We have also compared the effective moment of inertia of the
compound nucleus for the reaction systems that formed the similar
$^{248}\texttt{Cf}$ compound nucleus. we have obtained the average
errors of the calculated values for the
${^{12}\texttt{C}+^{236}\texttt{U}}$ and
${^{16}\texttt{O}+^{232}\texttt{Th}}$ reaction systems, over the
energy range for the reactions mentioned above about $6\%$ and
$<3\%$ in comparison with those of the
${^{11}\texttt{B}+^{237}\texttt{Np}}$ reaction system. It has been
shown in many instances the the observed anisotropies in fission
fragment angular distributions measured in heavy ion induced fission
on actinide targets deviate from the predictions of SSPSM. It is
generally viewed that this observed fission events consist of an
admixture of events of two types: compound nucleus fission(CNF), and
non compound nucleus fission(NCNF). The probability of non compound
nucleus of fission, $P_{NCNF}$  is given by
$P_{NCNF}(I)=\exp[{\frac{-0.5B_{f}(I,~ K=0)}{T}}]$, where $B_{f}(I,
K)$ and T are fission barrier height and nuclear temperature,
respectively[13]. For the reaction systems where the entrance
channel mass asymmetry $\alpha=(\frac{A_{T}-A_{P}}{A_{T}+A_{P}})$ is
greater than the Businaro-Gallone(BG) mass asymmetry $\alpha_{BG}$,
the measured anisotropies are in agreement with SSPSM predictions
and NCNF is absent. On the other hand, it is observed an anomalous
behavior in fission fragment angular distribution for the systems
with $\alpha$, smaller than $\alpha_{BG}$. While the
${^{11}\texttt{B}+^{237}\texttt{Np}}$ and
${^{12}\texttt{C}+^{236}\texttt{U}}$ reaction systems have a normal
behavior in fission anisotropies, the contribution of NCNF for the
${^{16}\texttt{O}+^{232}\texttt{Th}}$ reaction system is less than
$<10\%$ over the present energy range of investigation. The effect
of this contribution NCNF for the above reaction is the lower
average error, that is $<3\%$.

In conclusion, the calculation of the values of effective moment of
inertia using the experimental data for fission fragment angular
distribution as well as the prediction of the statistical
models is a novel method, which has been carried out in this work for the first time .\\

In this work, $\Im_{eff}$  is calculated in terms of emitting one
and two neutrons and compared with the case of ignoring the neutron
emission. It was noted that the values of $\Im_{eff}$ , due to
the neutron evaporation, increases, usually less than 10\%.\\
 Considering the level density parameter as
$\emph{a}=\frac{A_{C. N.}}{8}, \frac{A_{C. N.}}{10}, \frac{A_{C.
N.}}{11}$, rather than $\emph{a}=0.094A_{C. N.}$, the quantity for
effective moment of inertia of the compound nucleus varies at the
most 6 to 7 \%. Hence, this quantity is not sensitive to the level
density parameter selected in the computation.\\
 Overall, as experimental values of fission fragment
angular distribution are used in this method, we can conclude that
fission fragment angular distribution has been successful to
calculate the effective moment of inertia of the compound nucleus.

\end{document}